\documentclass[10pt]{article}

\usepackage{grffile}

\usepackage[small]{caption}
\usepackage{graphicx}

\usepackage{amsmath,amsfonts,amssymb}
\usepackage{amsthm}
\usepackage{color}

\usepackage[letterpaper,textheight=9in,textwidth=7in]{geometry}

\usepackage[skip=10pt,font=footnotesize]{caption}
\graphicspath{{Figures/}}

\newtheorem{Lemma}{Lemma}[section]
\newtheorem{Theorem}{Theorem}
\newtheorem{Proposition}[Lemma]{Proposition}

\newtheorem{Remark}[Lemma]{Remark}

\newenvironment{Proof}%
 {\begin{trivlist} \item[]{\bf Proof. }}%
 {\hspace*{\fill}$\rule{.4\baselineskip}{.4\baselineskip}$\end{trivlist}}

 {\begin{trivlist}\item[]\textbf{Acknowledgments.}}{\end{trivlist}}

\makeatletter\@addtoreset{figure}{section}\makeatother

\makeatletter \@addtoreset{equation}{section} \makeatother

\def\Xint#1{\mathchoice
{\XXint\displaystyle\textstyle{#1}}%
{\XXint\textstyle\scriptstyle{#1}}%
{\XXint\scriptstyle\scriptscriptstyle{#1}}%
{\XXint\scriptscriptstyle\scriptscriptstyle{#1}}%
\!\int}
\def\XXint#1#2#3{{\setbox0=\hbox{$#1{#2#3}{\int}$ }
\vcenter{\hbox{$#2#3$ }}\kern-.6\wd0}}

\def\dashint{\Xint-}
\newcommand{\R}{\mathbb{R}}

\newcommand{\Z}{\mathbb{Z}}

\newcommand{\ba}{\begin{align}}
\newcommand{\ea}{\end{align}}

\newcommand{\rmi}{\mathrm{i}}

\newcommand{\rmd}{\mathrm{d}}

\newcommand{\rme}{\mathrm{e}}

\newcommand{\rmO}{\mathcal{O}}

\newcommand{\eps}{{\varepsilon}}
\newcommand{\abs}[1]{\left | #1 \right |}
\newcommand{\norm}[1]{ \| #1  \|}
\newcommand{\bke}[1]{\left ( #1 \right )}

\begin{document}
\begin{center}
{\fontsize{16}{16}\fontfamily{cmr}\fontseries{b}\selectfont{Global phase diagrams of run-and-tumble dynamics: equidistribution, waves, and blowup}}\\[0.2in]
Kyungkeun Kang$\,^1$, Arnd Scheel$\,^2$, and Angela Stevens$\,^3$\\
\textit{\footnotesize 
$\,^1$Yonsei University, Department of Mathematics, 50 Yonsei-ro, Seoul 130-722, Republic of Korea \\
$\,^2$University of Minnesota, School of Mathematics,   206 Church St. S.E., Minneapolis, MN 55455, USA\\
$\,^3$University of M\"unster (WWU), Applied Mathematics, Einsteinstr. 62,
D-48149 M\"unster, Germany}
\date{\small \today} 
\end{center}

\abstract{
\setlength{\parindent}{0pt}
For spatially one-dimensional run-and-tumble
dynamics with mass conservation we develop a coarse phase diagram, that discriminates between global decay to equidistributed 
constant states, existence of spatially non-trivial waves, and
finite time blowup of solutions.  Motivated by counter-migrating
ripples of high and low population density and fruiting body formation in  myxobacteria,
we identify phase boundaries as particular critical tumbling dynamics that allow for switching between these spatio-temporal phases upon slight changes in mass densities or parameter values. 
}

\setlength{\parskip}{4pt}
\setlength{\parindent}{0pt}
\section{Introduction}

We study simple systems for run-and-tumble processes, modeled by a coupled pair of evolution equations for mass densities of left- and right-moving populations on the real line,
\begin{align}
 u_t&=\eps^2 u_{xx}+u_x+f(u,v),\notag\\
 v_t&=\eps^2 v_{xx}+-v_x - f(u,v).\label{e:cpw}
\end{align}
Here, $u(t,x)$ and $v(t,x)$ encode the densities of left- and right-traveling agents, respectively,  $f(u,v)$ is the total tumbling rate, encoding the frequency of changes in direction, the advection terms $u_x$ and $-v_x$ encode the constant speed of motion to the left and right, respectively, and $\eps^2 \partial_{xx}$, $\eps^2\geq 0$,  reflects possibly additional diffusive motion. The reversal rate is assumed to respect the reflection symmetry $u\leftrightarrow v$, $x\leftrightarrow -x$, that is, $f(u,v)=-f(v,u)$. We will consider \eqref{e:cpw} with periodic boundary conditions on an interval $x\in(0,L)$. 

While one can conceive many circumstances, where such simple models describe the basic dynamic mechanism, our inspiration stems from experimental observations of counter migrating rippling patterns in populations of myxobacteria, subject to 
environmental stress, especially starvation conditions. Myxobacteria have been observed to move at constant speed but spontaneously change direction, triggered in particular by the transmission of a so-called C-signal upon end-to-end contact. In addition to the emergence of rippling patterns, rather than simple uniform density distribution, one can observe the formation of fruiting bodies, that is, of three-dimensional
structures, after mass is concentrated in spatially localized regions. In our point of view, the onset of this process is reflected by blow-up in finite time, similar to the classical case of {\sl{Dictyostelium discoideum}} and
chemotaxis \cite{Nanjundiah, Childress-Percus, Jaeger-Luckhaus}; see for instance the analysis and simulations of models of type \eqref{e:cpw} in \cite{Lutscher-Stevens}.

In this context, more specific choices for $f$ are of the form
\begin{equation}\label{e:rate}
 f(u,v)=-u g(v)+v g(u),
\end{equation}
where now $g$ encodes a rate of reversal, depending on the rate of contact with counter-propagating agents. Our main  focus here is on monotone functions $g$, describing a likelihood of reversal that increases with the density of counter-propagating agents. Simple examples of interest to us are functions
\begin{equation}
 g(\rho)=\mu+\frac{\rho^p}{1+\gamma \rho^q}\label{e:g}.
\end{equation}
Here, $\mu$ reflects a ``spontaneous'' reversal rate, $p$ a rate of increase of the reversal rate with the frequency of contact for small population densities, $p-q$ the of rate increase (or decrease when $p<q$) for large population densities, and $\gamma$ encodes the transition density, or a saturation level, when reversal rates cease to increase at the same rate as for small densities. Notably, the case $\gamma=0$,
\begin{equation}
 g(\rho)=\mu+\rho^p\label{e:g0}
\end{equation}
corresponds to the absence of saturation and power-law increase of reversal rates for all densities. We are especially interested in critical reversal rates - as in \cite{Lutscher-Stevens}, such that a switch between rippling patterns and finite time blowup is
possible. After long rippling phases often fruiting bodies start to develop in 
experiments \cite{ShKai,Reich}.

\begin{figure}[h]
        \centering
        \includegraphics[width=0.245\linewidth]{eq_0.3.pdf}\hfill\includegraphics[width=0.24\linewidth]{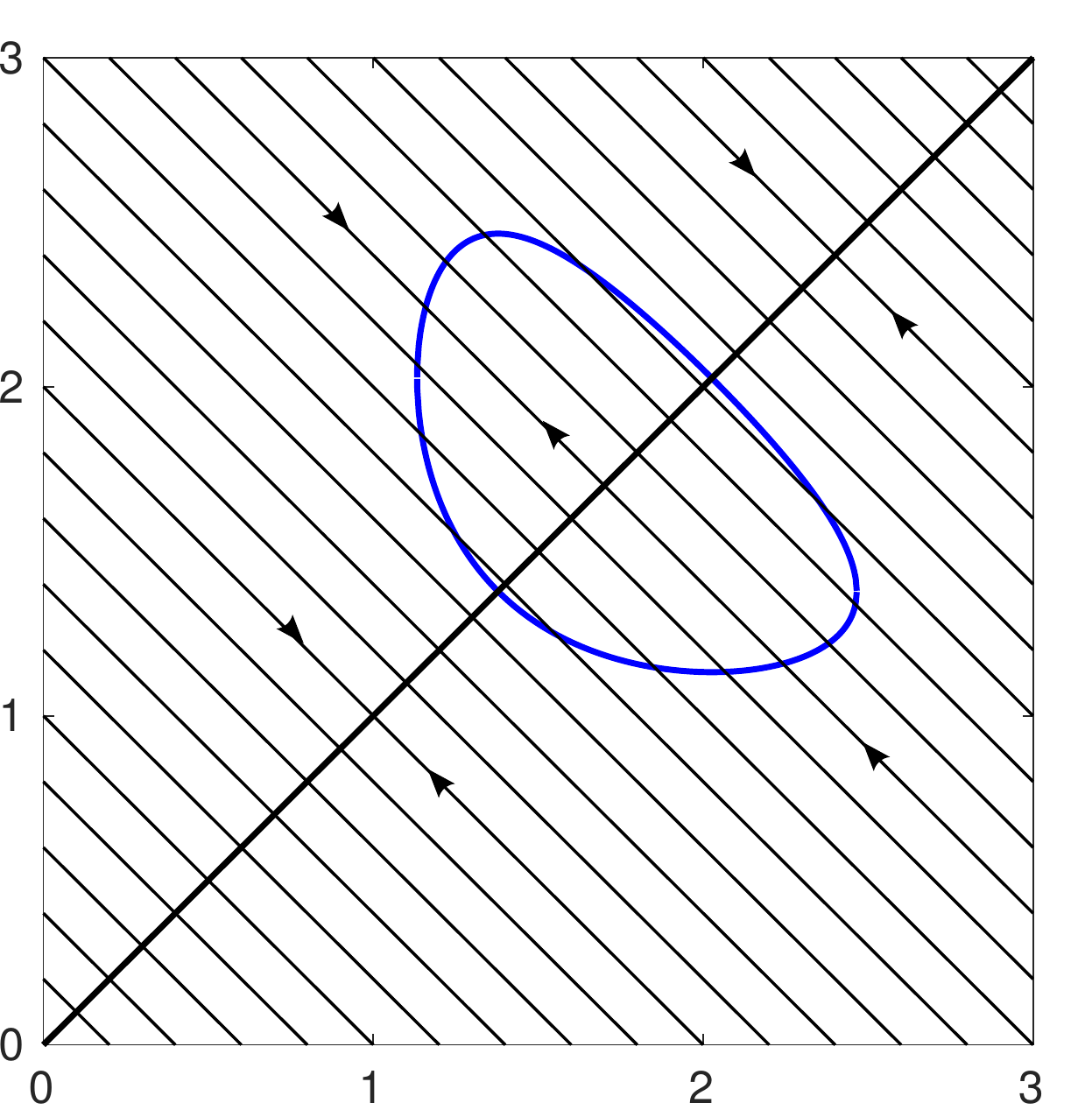}\hfill\includegraphics[width=0.245\linewidth]{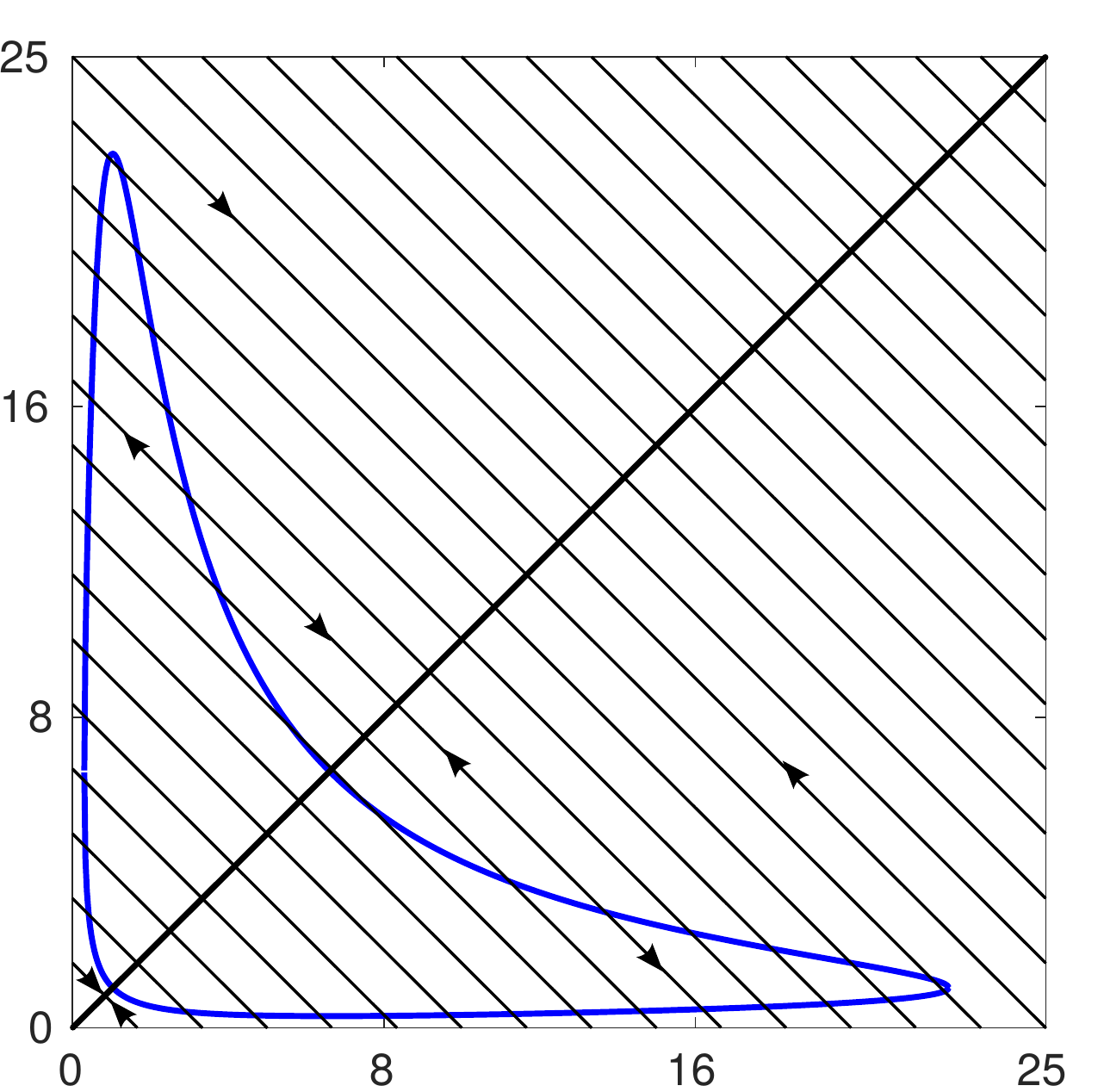}\hfill\includegraphics[width=0.23\linewidth]{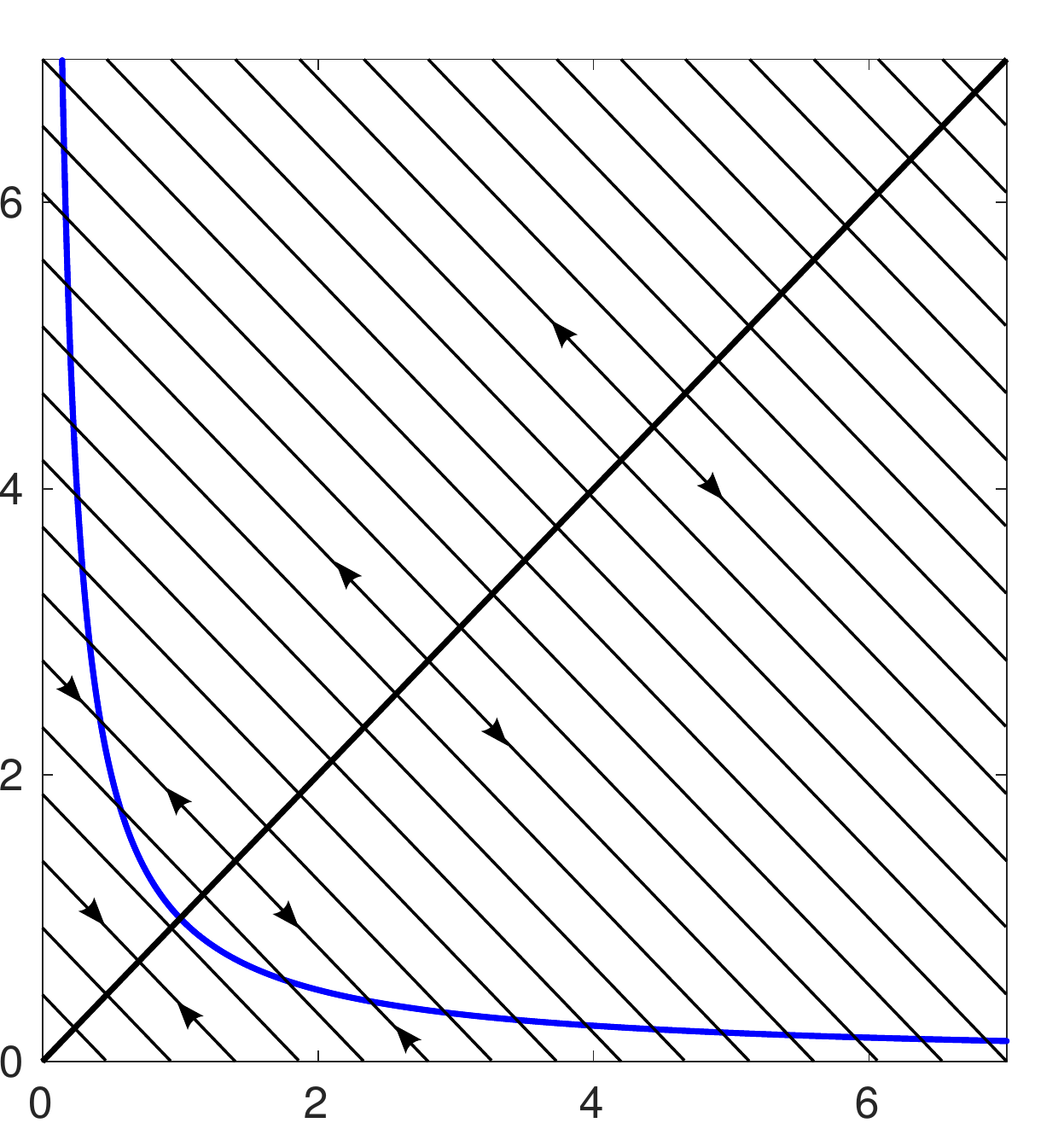}
        \caption{Dynamics of tumbling kinetics \eqref{e:cpwo}  in the $(u,v)$-phase 
plane, using \eqref{e:g} with parameter values $\mu=1,p=q=2$, and $\gamma=0.3,0.115,0.021,0$ from left 
to right. No asymmetric equilibria, $\gamma\in\Gamma_0$ on the left, bounded sets of asymmetric equilibria, $\gamma\in\Gamma_\mathrm{I}$ (center left and center right), and an unbounded set of asymmetric equilibria, $\gamma\in\Gamma_\infty$ (right). }\label{f:eq}
\end{figure}

\paragraph{Asymmetric ODE equilibria --- qualitative analysis.}
A study of dynamics of \eqref{e:cpw} would usually start with the behavior of spatially homogeneous, $x$-independent solutions,
\begin{align}
 u_t&=f(u,v),\notag\\
 v_t&= - f(u,v).\label{e:cpwo}
\end{align}
After exploiting the fact that mass $u+v$ is conserved, this 
ODE in the $(u,v)$-plane can be understood completely.
One therefore finds a one-dimensional equation with monotone dynamics, 
fully determined by the equilibria, where $f(u,v)=0$. By the built-in reflection symmetry, $f(u,u)=0$, that is, equidistributed densities, in space $x$ and across left- and right-moving populations, are equilibria. We refer to such equilibria as equidistributed, or symmetric. More interestingly, there may exist equilibria $f(u,v)=0$, $u\neq v$, which we will refer to as \emph{asymmetric equilibria}. Such asymmetric equilibria typically come in families, parameterized for instance by the total mass $u+v$. Figure \ref{f:eq} shows the dynamics for several typical parameter values. 
A first simple qualitative distinction between different phase diagrams would characterize the set of asymmetric equilibria as empty, bounded, or unbounded:
\begin{itemize}
 \item $\mathbf{\Gamma_0}$\textbf{ --- symmetric equilibria:} no asymmetric, only symmetric equilibria exist;
 \item $\mathbf{\Gamma_\mathrm{I}}$\textbf{ --- bounded asymmetric equilibria:} the set of asymmetric equilibria is bounded, non-empty;
 \item $\mathbf{\Gamma_\infty}$\textbf{ --- unbounded:} the set of asymmetric equilibria is unbounded. 
\end{itemize}
In the simple example $p=q=2$, $\mu=1$, elementary algebra shows that the sets $\Gamma_j$ correspond to intervals of $\gamma$-values, with 
\[
 \Gamma_0=[1/8,\infty),\qquad \Gamma_{\mathrm{I}}=(0,1/8),\qquad \Gamma_\infty=\{0\};
\]
see Figure \ref{f:eq} and \cite{scst}. Transitions between $\Gamma_1$ and $\Gamma_\infty$ in terms of $p,q$ are discussed in Proposition \ref{p:hom}.

\paragraph{PDE dynamics versus ODE equilibria --- main results.}
We establish a coarse ``phase diagram'' for the dynamics of \eqref{e:cpw}. 
Roughly speaking, a number of results are presented that corroborate the conjecture that the qualitative dynamics are determined by the phase regions $\Gamma_j$. To be somewhat more precise, we distinguish PDE dynamics qualitatively according to 
\begin{itemize}
 \item $\mathbf{\Upsilon_0}$\textbf{ --- equidistribution:}  global decay to equidistributed constant states;
 \item $\mathbf{\Upsilon_\mathrm{I}}$\textbf{ --- ripples \& waves:} global existence with uniform bounds \& existence of spatially non-trivial waves;
 \item $\mathbf{\Upsilon_\infty}$\textbf{ --- blowup:}  existence of finite-time blowup solutions.
\end{itemize}
In this sense, we show that, with some caveats and details, 
\begin{equation}\label{e:phd}
\textbf{ODE phase = PDE phase:}\hspace*{.7in} \Gamma_j=\Upsilon_j, \qquad\qquad j\in\{0,\mathrm{I},\infty\}.
\end{equation}
In other words, existence and boundedness of asymmetric ODE equilibria completely determines existence and boundedness of (non-trivial) PDE dynamics. Pushing the analogy to phrasing, the phase portrait of the ODE determines the rough ``phase diagram'' of the PDE dynamics. We expect and hope that many of the assumptions made below can be considerably weakened and a description of the phase diagram \eqref{e:phd} can be made more general and precise. 

In somewhat more detail, we start by basic existence theory, both in case $\eps=0$ and $\eps>0$. We show that nontrivial wave patterns exist in regions $\Gamma_\mathrm{I}\cup \Gamma_\infty$ for $\eps=0$, relying on an explicit construction from \cite{scst} in the case $\eps=0$ and a bifurcation argument in the case $\eps>0$. We establish global existence in $\Gamma_0\cup\Gamma_\mathrm{I}$ for $\eps\geq 0$, and equidistribution in $\Gamma_0$ for $\eps>0$ using a Lyapunov function. In the case $\eps=0$, we show conditional equidistribution, that is, equidistribution along convergent subsequences, and equidistribution for small data. In order to characterize $\Upsilon_\infty$, we construct self-similar blowup solutions for $p=2,\mu=0,\gamma=0,\eps=0$, establishing that $\Gamma_\infty=\Upsilon_\infty$ in this specific case. On the other hand, we show that $\Upsilon_\infty=\emptyset$ when $\eps>0$, and that solutions are in fact a priori bounded in time. 

\paragraph{From turning rates to function: a brief synopsis.} A coarse summary of our results, in light of the application to the collective behavior of populations with simple run-and-tumble laws, could focus on the ability of systems to collectively adapt and self-organize by tuning parameters only slightly, or by reacting to slight changes in environmental conditions. In regard to parameter changes, our contribution identifies the effect of parameters on the set of asymmetric equilibria as the key ingredient. Parameter value changes near $\gamma = 1/8$  allow for switching between rippling and equidistribution. Parameter value changes near $\gamma=0$ can lead the transition from rippling patterns to high-density clusters and subsequent fruiting body formation, identified here as finite-time blowup of density profiles. We do not pursue here the possibility of macroscopic changes in behavior due to changes in initial population configuration, such as the total mass at initial time. From our results one can conclude quite readily equidistribution in regions $\Gamma_\mathrm{I}$ and $\Gamma_\infty$, for initial conditions of sufficiently low mass (and even for high total mass in case $\Gamma_\mathrm{I}$). Similarly, rippling patterns still exist and may be stable in region $\Gamma_\infty$. It would clearly be interesting to discriminate these finer properties within the phase diagram developed here. 

\paragraph{Outline.} We discuss basic properties of \eqref{e:cpw} in Section \ref{s:2} and show the existence of nontrivial ripple patterns in Section \ref{s:hom}, thus characterizing the region $\Upsilon_1$. Section \ref{s:4} contains results on equidistribution, region $\Upsilon_0$, and Section \ref{s:5} states a blowup result pertaining to $\Upsilon_\infty$, $\eps=0$. Section \ref{s:6} contains a global existence result with uniform a priori bounds in the case $\eps>0$, showing that diffusion does not even allow for growth of solutions and thus also not for their blowup. We conclude with a brief discussion. 

\paragraph{Acknowledgments.} The research of KK was partially supported by NRF-2017R1A2B4006484. ASc. gratefully acknowledges support through NSF grant DMS-1612441 and through  a Humboldt Research Award. KK and ASc. gratefully acknowledge generous hospitality at WWU M\"unster during a research stay when this research was initiated.

\section{Local existence and flow properties}\label{s:2}

We briefly review basic existence theory for the class of equations considered here. Recall that we focus on periodic boundary conditions on the interval $x\in (0,1)$ with identification $0\sim 1$, denoted as $S^1=\R/\Z$. Other interval lengths can be obtained by scaling $t,x$. Possibly more relevant boundary conditions such as reflection boundary conditions $u=v$ at $x=0,1$ can be reduced to the case of periodic boundary conditions on $x\in(-1,1)$ by reflection, $u(-x)=v(x),\ x\in (-1,1)$, Throughout we will assume that $f$ is sufficiently smooth, in particular locally Lipshitz,
c.f. also \cite{Lutscher}.

\paragraph{Pure transport: $\eps=0$.} In the absence of diffusion, one can find solutions almost explicitly. 
Inverting the transport operators $\partial_t\pm\partial_x$ by integrating along characteristics, one readily obtains the Picard-type integral formula
\begin{align}
 u(t,x-t)=u_0(x)+\int_0^t f(u(\tau,x-\tau),v(\tau,x-\tau))\rmd\tau,\notag\\
 v(t,x+t)=v_0(x)-\int_0^t f(u(\tau,x+\tau),v(\tau,x+\tau))\rmd\tau,\label{e:voc}.
\end{align}

\begin{Theorem}[Local existence, $\eps=0$]\label{t:loc}
Let $\Sigma:=\{|t|\leq T, \ x\in S^1\}$.
 For all $(u_0,v_0)\in L^\infty$, there exists $T>0$ and functions $u(t,x),v(t,x)\in L^\infty(\Sigma)$, for $(t,x)\in \Sigma$,  
which are unique solutions to \eqref{e:voc}. Moreover $u(t,x),v(t,x)$ are weak solutions to \eqref{e:cpw}, with $\eps=0$, that is, 
 \begin{align*}
  \int_{x=0}^1\int_{t=-T}^T \left(u(\varphi_t-\varphi_x)+f(u,v)\varphi\right)\rmd t\rmd x=0,\\
  \int_{x=0}^1\int_{t=-T}^T \left(v(\psi_t+\psi_x)-f(u,v)\psi\right)\rmd t\rmd x=0,
 \end{align*}
 for all $\varphi,\psi\in C^\infty_0(\Sigma)$. Further, the solution $(u,v)\in L^\infty(\Sigma)$ depends continuously on $(u_0,v_0)\in S^1$. In addition, $u(t,\cdot-t)$ and $v(t,\cdot+t)$ are continuous in $t$  as functions in $L^\infty(S^1)$.  
\end{Theorem}

\begin{Proof}
 Since $f$ is locally Lipshitz, it defines a locally Lipshitz superposition operator on $L^\infty(\Sigma)$. The shear transformations $u(t,x)\mapsto u(t,x\pm t)$ are bounded operators, as well. Integration defines a bounded linear map with small norm when $T\ll 1$, such that, in complete analogy to the existence theory for ODEs, we obtain local existence and uniqueness, together with continuous dependence on initial conditions. Noticing that the right-hand side possesses a weak derivative in $t$, one finds that $(\partial_t-\partial_x)u$ exists in the weak sense, establishing that our Duhamel formula indeed yields weak solutions.
\end{Proof}
\begin{Remark}\label{r:lp}
For $f$ globally Lipshitz, the same construction yields global solutions for initial data in $L^p(S^1)$, $1\leq p<\infty$ depending continuously on the initial data in $L^p$. Since the shift is continuous on $L^p$, one also obtains continuity in time of $u(t,\cdot)$ and $v(t,\cdot)$.  More directly, we see immediately that global existence also holds in $C^0$ with equivalent bounds.
%
%
%

\end{Remark}

\paragraph{Transport and diffusion: $\eps>0$.}
Adding diffusion, the problem turns into a semilinear parabolic equation with solutions that are smooth for small positive times.

\begin{Theorem}[Local existence, $\eps>0$]\label{t:locd}
 For all $(u_0,v_0)\in L^\infty$, there exists $T>0$ and functions $u(t,x),v(t,x)\in C^\infty((0,T]\times S^1)$, which are unique classical solutions to \eqref{e:voc} with $\eps>0$, such that $u(t,\cdot), v(t,\cdot)$ converge to $u_0,v_0$ weakly. Here weak convergence is understood as convergence in $L^\infty$ after regularizing with $(1-\partial_{xx})^{-1}$. Moreover, the solution $(u,v)\in C^k(\Sigma)$ depends continuously on $(u_0,v_0)\in S^1$ for any  $0<t\leq T$ and any $k<\infty$. 
\end{Theorem}
\begin{Proof}
 The equation now fits into the framework of analytic semigroups with smooth nonlinearities. Indeed, smoothness of $f$ implies smoothness of the superposition operator in $L^\infty$, and the linear part is given by a sectorial operator. One then obtains a unique local solution, using for instance \cite[Thm 7.1.2]{lun}. Bootstrapping, that is, using that the nonlinearity also defines smooth superposition operators on spaces of smooth functions gives a solution in $C^k$ and higher differentiability in time. 
\end{Proof}

\begin{Remark}[Global existence]
  We note that in both cases, $\eps=0$ and $\eps>0$, the time of existence $T>0$ depends only on $|(u,v)|_\infty$, such that the existence of global solutions 
would follow from 
  a priori bounds for solutions in $L^\infty$.
\end{Remark}

\section{Space-homogeneous equilibria and ripples}\label{s:hom}

Now we characterize $\Gamma_\mathrm{I}$ and $\Upsilon_\mathrm{I}$. We first discuss boundedness of the branch of equilibria, discriminating $\Gamma_\mathrm{I}$ from $\Gamma_\infty$. We then establish the existence of nontrivial wave solutions in $\Upsilon_\mathrm{I}\cup\Upsilon_\infty$.

\begin{Proposition}[Asymmetric equilibria near infinity]\label{p:hom}
Consider turning rates $g(v)=\mu+\frac{v^p}{1+\gamma v^q}$ with $p,q\geq 1$, $\mu,\gamma>0$. Then the set of spatially constant, asymmetric equilibria $(u,v)\in\R^2$ of (\ref{e:cpw}), solutions to $ug(v)=vg(u)$, is  unbounded for $p>q+1$, and for $p=q+1$ whenever the function $\Pi(u)=\mu - \gamma u + \mu \gamma^2 u^q$  possesses a simple real root in $u>0$. The set of asymmetric equilibria is bounded when $p<q+1$ or when $p=q+1$ and $\Pi(u)$ is strictly positive in $u\geq 0$.  
\end{Proposition}
\begin{Proof}
 We regularize the equation $f(u,v)=0$ by multiplying with $(1+\gamma u^q)(1+\gamma v^q)$, obtaining a polynomial expression in the variables $u,v$ and their $p$'th and $q$'th powers,
 \[
  v(\mu+\mu\gamma u^q + u^p)(1+\gamma v^q)=u(\mu+\gamma\mu v^q+v^p)(1+\gamma u^q).
 \]
The power-law behavior at infinity allows us to study zeros near infinity using inversion, setting $u=1/a,v=1/b$, thus obtaining a polynomial equation in these new variables and their $p$'th and $q$'th powers. We are interested in zeros to this new equation in $(u,b)$ or $(a,b)$ near $b=0$, $u\geq0$, and near $a=b=0$, respectively. 

Consider first the case $p< q+1$. A short calculation, using for instance some version of the Newton polygon, gives the leading order term $\mu+\gamma\mu u^q+u^p$ at $b=0$, which is strictly positive. We can therefore exclude zeros in a neighborhood of $v=\infty$, $u\geq 0$. Near $a=b=0$, one obtains at leading order $a^{q+1-p}-b^{q+1-p}=0$, with a  solution $\alpha=\beta$ in the new variables $\alpha=a^{q+1-p}, \beta=b^{q+1-p}$. Exploiting that error terms are superlinear in $\alpha,\beta$, the implicit function theorem guarantees
that this solution is unique. In summary, there are no nontrivial solutions near infinity. 

In the case $p>q+1$, one proceeds in a similar fashion and finds an equation $\mu b=u^{p-q-1}$ at leading order. Continuing with the implicit function theorem yields the nontrivial solution $v\sim\mu u^{q+1-p}$. 

In the case $p=q+1$, one again establishes that there are no nontrivial solutions near $a=b=0$, and finds the polynomial 
$\mu -\gamma\mu+\mu \gamma^2u^q$ at $b=0$. 

\end{Proof}

We next show that there exist non-trivial ripple patterns, i.e. counter migrating
traveling waves,  in  $\Gamma_\mathrm{I}\cup\Gamma_\infty$.

\begin{Lemma}[Rippling patterns, $\eps=0$]\label{l:ripple}
 Suppose there exist $u_*>v_*$ such that $f(u,v)=0$. Let $P_u,P_v\subset S^1$ be measurable and define
 \[
  u(t,x)=\left\{\begin{array}{ll} u_*,& x+t\in P_u,\\ v_*, & x+t\not\in P_u,\end{array}\right.\qquad v(t,x)=\left\{\begin{array}{ll} u_*,& x+t\in P_v,\\ v_*, & x+t\not\in P_v.\end{array}\right.
 \]
Then $u(t,x)$ is a weak solution to \eqref{e:cpw} in the sense of Theorem \ref{t:loc}. 
\end{Lemma}

\begin{Proof}
 Since $f(u,v)=0$ for $u,v\in\{u_*,v_*\}$, $f(u(t,x),v(t,x))\equiv 0$. The simple translation of initial conditions therefore gives a weak solution to the left- and right-shift equation for $u$ and $v$, respectively.
\end{Proof}

Combining Proposition \ref{p:hom} and Lemma \ref{l:ripple}, we see that ripples exist for $\gamma<1/8$. Moreover, the set of ripples is unbounded when $p>q+1$. 

In the case $\eps>0$, most of the ripple patterns will not persist. We refer however to \cite{fns}, where stationary, spatially localized clusters and gaps, as well as interfaces are constructed for $\gamma<1/12$, $\eps=1$. Such solutions appear to be dynamically stable for certain parameter regimes. Roughly speaking, regions of high densities of $u$ and $v$ are bounded to the left by regions of  even higher concentrations of $v$. This self-organized inward moving ``barrier'' prevents agents from escaping the region of high densities. To the right, regions of high density are analogously contained by barriers consisting of high $u$-concentrations. Traveling waves exist for larger values $\gamma<1/8$, regardless of $\eps$; see again \cite{Lutscher-Stevens, fns}, the earlier work \cite{fuhrmann} for such models applied to the dynamics
of the cellular cytoskeleton, and \cite{Freistuehler-Fuhrmann}.

\section{Global existence for bounded asymmetric states and relaxation to equidistribution}\label{s:4}
We place ourselves in the complement of region $\Upsilon_\infty$, where asymmetric states exist in a bounded region, only. More precisely, we assume  that $f(u,v)>0$ in $|(u,v)|\geq R$ for some $R>0$.

\begin{Theorem}[Global existence and uniform bounds, $\eps\geq 0$]\label{t:glob}
 Assuming boundedness of the set of asymmetric equilibria, the unique local solution in Theorem \ref{t:loc} can be extended globally in time $t\geq 0$. Moreover, $\sup_t \|(u,v)(t,\cdot)\|_\infty\leq \max\{R,\|(u,v)(0,\cdot)\|_\infty\}$, where we used the box norm $\|(u,v)\|_{\R^2}=\max\{|u|,|v|\}$.
\end{Theorem}

\begin{Proof}
This result was established in \cite{Hillen97} for $\eps=0$ and more general boundary conditions. We include a short proof. We start with the case $\eps=0$ and use a weak maximum principle for the evolution along characteristics. Suppose first that the essential supremum of $(u,v)$ at time $0$ is less than $R$ and that  $t_0$ is the first time where the essential supremum is $R$. Let $x_0$ be a point where $u(t_0,x_0)\geq R-\delta$, $v(t_0,x_0)\leq R$, such that $f<0$ at $t_0,x_0$.  Continuity of $u(t,\cdot -t)$ in $t$ and the fact that weak derivatives exist along characteristics then imply that $\sup_\mathrm{ess} |(u,v)|\geq R-\delta+\delta'$ for $t=t_0-\tau$, for some $\delta',\tau>0$ independent of $\delta$. Letting $\delta$ go to zero then proves the claim. For $\eps>0$, we can use the parabolic maximum principle and the fact that the rectangles $[0,r]^2$ are strictly forward invariant for the ODE when $f>R$; see for instance \cite{smoller}
\end{Proof}
As a consequence, we may modify the nonlinearity $f$ outside of a box of size $R$ such that the modified nonlinearity is globally Lipshitz, not affecting the dynamics for initial data within a ball of size $R$ in $L^\infty$. The flow restricted to such initial conditions is then continuous in $L^p$, $1\leq p<\infty$, even in the case $\eps=0$. 
\begin{Remark}[Attractor]
 Refining the argument, one can show that $\limsup_t \|(u,v)(t,\cdot)\|_\infty \leq R$, establishing the existence of an absorbing set. One would then define a global attractor as the $\omega$-limit set of a ball of size $R$; see for instance \cite{attr,attr2}. In the case $\eps\geq 0$, one would readily conclude from a priori bounds and smoothing properties of the flow that this global attractor is compact and finite-dimensional. In the case $\eps=0$, any characterization of the attractor appears to be difficult. It is for instance not a priori clear that the $\omega$-limit set of individual trajectories is always nonempty. In the opposite direction, this $\omega$-limit sets would contain the plethora of wave patterns constructed in Lemma \ref{l:ripple}, suggesting that it may well not be compact. 
\end{Remark}

We next focus on the question of equidistribution. Having established that solutions stay bounded for all times when the set of asymmetric equilibria is bounded, and nontrivial rippling patterns exist whenever this set is nonempty, we will show that solutions converge to a symmetric, equidistributed state, when there do not exist asymmetric solutions. 

\begin{Theorem}[Equidistribution, $\eps>0$]\label{t:equi}
 Fix $\eps>0$ and suppose that there do not exist any asymmetric equilibria. More precisely, we assume that $f$, defined in Section \ref{s:hom}, satisfies $f(u,v)>0$.

 Then, for any initial condition $u_0,v_0\in L^\infty$, we have
 \[
  \lim_{t\to\infty} (u(t,\cdot),v(t,\cdot))=(m/2,m/2), 
 \]
where $m=\dashint (u+v)$, with convergence in $C^k(S^1)$ for any $k<\infty$.
\end{Theorem}
\begin{Proof}
 We compute the $L^2$-energy estimate
  \begin{align}
  \frac{1}{2}\frac{\rmd}{\rmd t} \int_x \left(u^2+v^2\right)&=-\eps^2\int \left(u_x^2+v_x^2\right)-\int_x (u-v)^2f(u,v),\label{e:p21d}
\end{align}
and conclude that the $L^2$-norm is a Lyapunov function. Since the forward orbit is precompact, we conclude that the $\omega$-limit set is non-empty and is contained in a level set of the Lyapunov function by LaSalle's invariance principle; see for instance \cite{attr2}. Since the energy therefore is non-decreasing on the $\omega$-limit set, the right-hand side of \eqref{e:p21d} vanishes. The first term on the right-hand side implies that therefore $u,v$ are constant and the second term implies $u=v$ on the $\omega$-limit set. It remains to show that the $\omega$-limit set consists of a single equilibrium. We therefore restrict the flow to the affine spaces $\dashint u+v=m$, which is flow-invariant by mass conservation. Within each of those spaces, equilibria are isolated, such that the (connected) $\omega$-limit set consists of a single equilibrium, only.
\end{Proof}

\begin{Theorem}[Conditional equidistribution, $\eps=0$]\label{t:equi1}
 Set $\eps=0$ and suppose that there do not exist any asymmetric equilibria. More precisely, we assume that $f$, defined in Section \ref{s:hom}, satisfies $f(u,v)>0$. 

Fix any initial condition $u_0,v_0\in L^\infty$. 
Then for any sequence $t_k\to\infty$, we have
 \[
  \lim_{t_k\to\infty} (u(t_k,\cdot),v(t_k,\cdot))=(m/2,m/2), \qquad m=\dashint (u+v),
 \]
in $L^p$, $1\leq p\leq \infty$, provided that the limit exists in this space.
\end{Theorem}
%

\begin{Proof}
From  the energy estimate
 \begin{align}
  \frac{1}{p}\frac{\rmd}{\rmd t} \int_x \left(u^p+v^p\right)&=-\int_x (u-v)\left(u^{p-1}-v^{p-1}\right)f(u,v),\label{e:p22}
\end{align}
where $u^q:=u|u|^{q-1}$, we see that $W[u,v]:=\|(u,v)\|_p$ is non-increasing and a continuous function on $L^p$. 
Continuity of the flow in $L^p$ therefore implies the statement as follows. Let $\Phi_t$ denote the continuous flow on $L^p$, and suppose $(u(t_k,\cdot),v(t_k,\cdot))\to  (u_\infty(\cdot),v_\infty(\cdot))$. 
Then we conclude that 
\[
(u(t_k+T,\cdot),v(t_k+T,\cdot))\to  \Phi_T(u_\infty(\cdot),v_\infty(\cdot))=:(\bar{u}_\infty(T,\cdot),\bar{v}_\infty(T,\cdot)).
 \] 
Continuity of $W$ implies that $W_k:=W(u(t_k,\cdot),v(t_k,\cdot))\searrow W_\infty= W(u(t_\infty,\cdot),v(t_\infty,\cdot))$, and  $W'_k:=W(u(t_k+T,\cdot),v(t_k+T,\cdot))\searrow W'_\infty= W(\Phi_T(u_\infty(\cdot),v_\infty(\cdot)))$. The fact that $W$ is non-increasing along solutions implies that $W_{k+\ell}\leq W_k'\leq W_k$,  for $\ell$ sufficiently large, such that the limits of $W_k$ and $W_k'$ coincide, proving that $W$ is constant on the trajectory $(\bar{u}_\infty(T,\cdot),\bar{v}_\infty(T,\cdot)$. Inspecting the $L^2$-energy estimate, we see that therefore $\bar{u}_\infty(T,x)=\bar{v}_\infty(T,x)$ almost everywhere, for all $T$. As a consequence, using the equation $(u_\infty+v_\infty)_x=0$,  together with $u_\infty(x)=v_\infty(x)$ and mass conservation gives $u_\infty(x)\equiv v_\infty(x)\equiv m/2$.
\end{Proof}

\begin{Proposition}[Equidistribution for small data: $\eps=0$]\label{p:equi}
 Let $\eps=0$ and suppose that $f(u,v)>0$. 
%

Then there exists $\delta= \delta(m)>0$ such that for any  initial conditions with $|(u_0-m/2,v_0-m/2)|_{H^{1}}<\delta$, we have 
  \[
  \lim_{t\to\infty} (u(t,\cdot),v(t,\cdot))=(m/2,m/2),
 \]
 in $H^{1}(S^1)$.
\end{Proposition}
\begin{Proof}
 We restrict the semiflow to constant mass $\dashint (u+v)=m$, and linearize at equilibria, finding the linear equation 
 \begin{equation}\label{e:ls}
  u_t=u_x-(u-v)R,\qquad v_t=-v_x+(u-v)R.
 \end{equation}
We claim that $\dashint(u_0+v_0)=0$ implies that $|u(t,\cdot)|_\infty+|v(t,\cdot)|_\infty\leq C\rme^{-\kappa t}$ for some $C,\kappa>0$. A simple contraction argument for the nonlinear equation, exploiting the fact that the nonlinearity defines a smooth superposition operator in $H^1$, then gives the result.

To obtain the linear estimate, we first notice that it is sufficient to obtain the estimate in $L^2$, since the linearization commutes with translations. Using Fourier transform, it is therefore sufficient to analyze the matrix multiplication operator $\hat{\mathcal{L}}(u,v)^T(k)=((\rmi k -1)u+v,u + (-\rmi k  -1)v)^T$, with $k\in (2\pi/L)\Z$. Eigenvalues of the matrix are $\lambda_\pm(k)=-1\pm \sqrt{1-k^2}$, with eigenvectors $(1,\rmi k - \sqrt{1-k^2})^T$ and $\rmi k - \sqrt{1-k^2},-1)^T$, respectively. In particular, the diagonalizing transformation is uniformly bounded for large $k$, eigenvalues are bounded away from the imaginary axis except for the eigenvalue $\lambda_+(0)$ associated with mass conservation. We can therefore view the operator as a direct sum of a skew adjoint operator, corresponding to values of $k$ with algebraically simple, purely imaginary eigenvalues, a self-adjoint operator, corresponding to values of $k$ with real eigenvalues, and a possibly finite dimensional part with an algebraically double eigenvalue $\lambda=-1/2$, when $k=1$. In the self-adjoint and finite-dimensional part, one readily finds exponential decay. In the infinite-dimensional, skew-symmetric part, one finds exponential decay from an $L^2$-estimate. This concludes the proof. 
\end{Proof}

\begin{Remark}[Unconditional equidistribution for $\eps=0$]
We remark the key  difference between Theorem \ref{t:equi}  and Theorem \ref{t:equi1} is the additional assumption of convergence of subsequences. In our proof, we could avoid this assumption if  the forward orbit were precompact, which would reduce the statement to more classical versions of LaSalle's invariance principle. Such compactness is unfortunately not easily obtained given the lack of smoothing of the linear equation, evident in particular in the region $\Upsilon_1$ where a plethora of ripple patterns exist. Alternatives to compactness are coercivity estimates of various kinds, as illustrated in Proposition \ref{p:equi}. In particular, one would like to conclude that $\int_0^\infty |u(t,\cdot)-v(t,\cdot)|_p^p<\infty$ for $p=1$, whereas the energy estimates presented here only give this estimate for $p=2$. 
\end{Remark}

\section{Blowup profiles: $\eps=0$}\label{s:5}
%
%

In this section, we address the case of the region $\Gamma_\infty$  where the set  of asymmetric equilibria is unbounded, for instance $\gamma=0$ or $p\geq q+1$. Since one expects blowup to be governed by the behavior at large amplitudes, we investigate scale-invariant turning rates, $\gamma=\mu=0$,
\[
 u_t=u_x-uv^p+vu^p,\qquad v_t=-v_x + uv^p-vu^p,
\]
and moreover focus on the case $p=2$. Our main result in this case establishes the existence of a family of solutions that blow up in finite time, that is, the $L^\infty$-norm of the solution diverges as $t\to T_*$.  In the terminology of a phase diagram of dynamics, this establishes that this particular element of $\Gamma_\infty$ belongs to $\Upsilon_\infty$.

\begin{Theorem}[Self-similar blowup]\label{t:blow}
For any $\delta>0$, there exists a family of initial conditions $u_0(x;M),v_0(x;M)$, $u_0(-x;M)=v_0(x;M)$, defined on $x\in(-\delta,\delta)$, bounded and smooth, parameterized by the value $u_0(0;M)=v_0(0;M)=M$, such that any initial condition that coincides with $u_0,v_0$ on $x\in(-\delta,\delta)$ will give rise to a solution $(u(t,x),v(t,x))$ that blows up in finite time, that is, there exists $T_*$ such that 
\[
 \sup_x(|u(t,x)|+|v(t,x)|)\nearrow \infty\ \mbox{ for } t\to T_*.
\]
\end{Theorem}
\begin{Proof}
 We construct self-similar solutions of the form 
 \[
  u(t,x)=(-t)^{-1/2}U(x/(-t)),\qquad v(t,x)=(-t)^{-1/2}V(x/(-t)),
 \]
defined on $t<0$, $|x/t|<1$. Clearly, taking initial conditions at time $t=-\delta$ on $(-\delta,\delta)$ will lead to blowup at $t=0$. Since the value of the solution in $|x/t|<1$ only depends on values in this sector, by construction of the solution via integration along characteristics, one can extend initial conditions arbitrarily outside of $(-\delta,\delta)$ without changing the solution within the cone $|x/t|<1$.

Self-similar solutions solve a non-autonomous ODEs w.r.t. $\xi=x/(-t)\in(-1,1)$,
namely 
\begin{align*}
 (-1+\xi)U'+\frac{1}{2}U&=-UV^2+VU^2,\\
 (1+\xi)V'+\frac{1}{2}V&=UV^2-VU^2.
 \end{align*}
We rescale the time interval, setting $\xi=\tanh(y)$, where now $y\in (-\infty,\infty)$ corresponds to $\xi\in (-1,1)$. This gives the (autonomous) system in 3-dimensional phase space,
\begin{align}
 U_y&=-(1+\xi)\left(-\frac{1}{2}U-UV^2+VU^2\right),\notag\\
 V_y&=(1-\xi)\left(-\frac{1}{2}V+UV^2-VU^2\right),\notag\\
 \xi_y&=1-\xi^2.\label{e:3d}
\end{align}
Our goal is to find bounded solutions to this equation. We next describe basic properties of the dynamics of this 3-dimensional ODE. First, we notice that the affine subspaces $\{\xi=-1\}$ and $\{\xi=+1\}$ are invariant\footnote{Note that $\xi=1$ corresponds to $x=-1$ at $t=-1$, that is, orientation of $\xi$ and $x$ are reversed by the transformation.}. For $\xi=1$, $V$ is constant in time and the $U$-dynamics are 
\[
 U'=-2U\left(-\frac{1}{2}-V^2+VU\right),
\]
with equilibria $U=0$ and $U=({V^2+\frac{1}{2}})\big / {V}$. Similarly, we find for the  dynamics in $\{\xi=-1\}$ that $U$ is constant and 
\[
 V'=2V\left(-\frac{1}{2}-U^2+VU\right),
\]
with equilibria $V=0$ and $V=({U^2+\frac{1}{2}})\big/{U}$. In particular, equilibria with $U,V\neq 0$ are stable in $\xi=1$ and unstable in $\xi=-1$; see Figure \ref{f:1} for phase portraits.
\begin{figure}
\centering
 \includegraphics[width=0.25\textwidth]{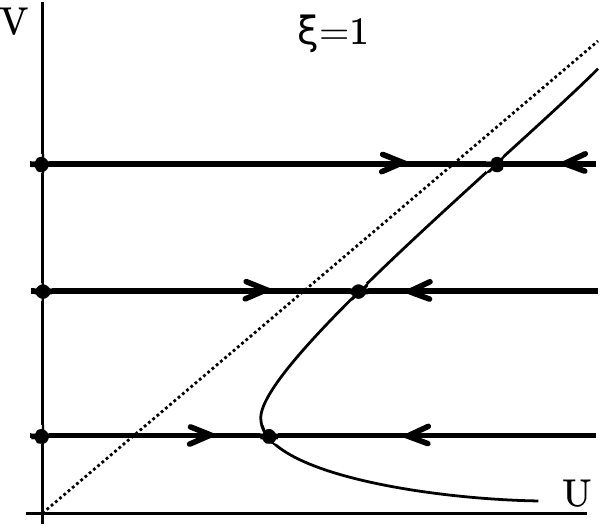}\hspace*{0.08\textwidth}\includegraphics[width=0.25\textwidth]{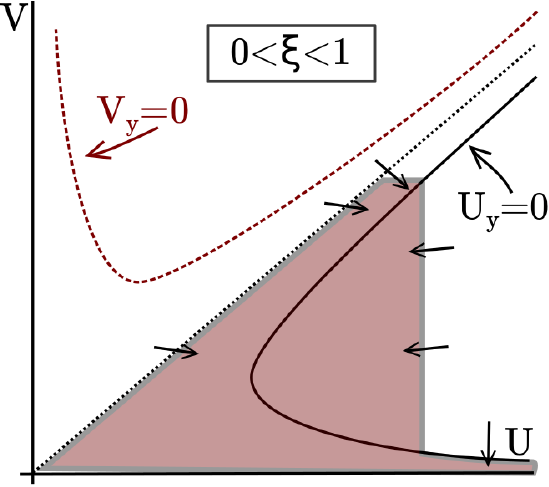}\hspace*{0.08\textwidth}\includegraphics[width=0.25\textwidth]{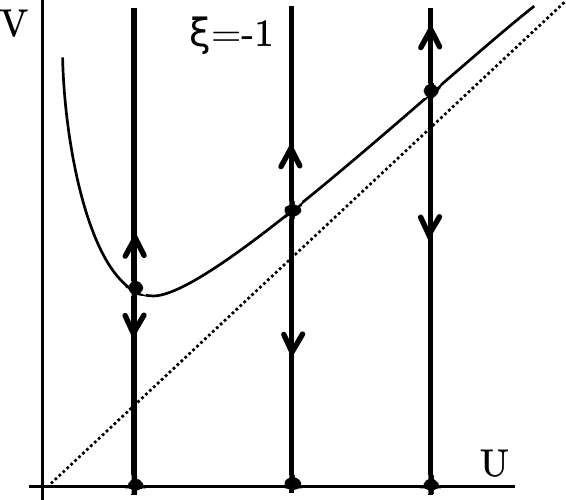}
 \caption{Phase portraits at $\xi=1,\xi=-1,$ and $0<\xi<1$ from left to right. }\label{f:1}
\end{figure}
Next, note that reflection symmetry of the original system translates into a reversibility of \eqref{e:3d}. More precisely, any solution $(u,v,\xi)(y)$ gives rise to a solution $(v,u,-\xi)(-y)$. In particular, any bounded solution $(u,v,\xi)(y)$ in $y\geq 0$ with $u=v,\xi=0$ at $y=0$ gives rise to a bounded solution on $y\in\R$. We may therefore employ a shooting strategy, taking an initial condition $U=V=m$, $\xi=0$, and solve forward in time $y$. Showing that this solution remains bounded for all $m$ then proves the theorem.

To start with, we clearly have $\xi>0$ for $y>0$ and $\xi\to 1$ as $y\to\infty$. 
Now, notice that the region $V\leq U$ is forward invariant in $\xi\geq 0$, that is, initial conditions in this region give rise to trajectories in this region. Indeed, we inspect the equation at the boundary, where $U=V>0$, and find 
\[
 (V-U)_y=(1-\xi)\frac{-1}{2}V+(1+\xi)\frac{-1}{2}V<0. 
\]
such that any solution on the boundary will enter the region; see Figure \ref{f:1}, middle panel, for an illustration of the invariant regions described here and in the following. 

Next, examine the curve $U=V+\frac{1}{2V}$, which corresponds to equilibria at $\xi=1$. On this curve, $U_y=0$ and $V_y=(1-\xi)V(-1-\frac{1}{4U^2})<0$.
Proceeding in a similar fashion, one verifies that the shaded region in Figure \ref{f:1} is forward invariant. As a consequence, we find that solutions are either bounded, or $U\to\infty$, $V\to 0$, for $y\to\infty$ or for $y\nearrow y^+$. 

It now remains to exclude the possibility $U\to\infty$, $V\to 0$. Inspecting the equation for $U$ in this region, we readily find that the maximal growth is achieved at $V=0$, such that $U'\leq 2U$, and $U\to\infty$ for $y\nearrow y^+<\infty$ is not possible. We therefore proceed via Poincar\'e inversion, setting $U=1/W$, which, regularizing with the Euler multiplier $W^2$, setting $\partial_\sigma=W^2\partial_y$,  yields
\begin{align*}
 {W}_\sigma&=(1+\xi)W^2\left(-\frac{1}{2} W+V-WV^2\right),\\
 {V}_\sigma&=(1-\xi)V\left(-\frac{1}{2}W^2-1+WV\right),\\
 {\xi}_\sigma&=W^2\left(1-\xi^2\right).
\end{align*}
We are now interested in the dynamics of this flow near $\xi=1,W=V=0$, and therefore substitute $\xi=1-\eta$, with new equation 
\begin{align}
 {W}_\sigma&=(2-\eta)W^2\left(-\frac{1}{2} W+V-WV^2\right),\notag\\
 {V}_\sigma&=\eta \left(-\frac{1}{2}VW^2-V+WV^2\right),\notag\\
 {\eta}_\sigma&=W^2\left(-2\eta+\eta^2\right).\label{e:inv}
\end{align}
In this system, we find equilibria $W=2V+\rmO(V^2)$, $\eta=0$, which correspond to the nontrivial equilibria of \eqref{e:3d} in $\xi=1$. Our goal is to prove that all trajectories with initial conditions in $0<\eta\leq 1$, $0<V<W/2$, converge to an equilibrium $W=2V+\rmO(V^2)>0$, thus establishing boundedness of $U=1/W$ as desired. Unfortunately, it is not straightforward to establish this fact for \eqref{e:inv}, in particular since the origin $W=V=\eta=0$ is a degenerate equilibrium. 

Recall that for $U\to\infty$, we have $\eta\to 0$ and $V\to 0$, such that any unbounded solution necessarily approaches the origin in \eqref{e:inv}. The remainder of the proof consists of describing the dynamics near the origin in \eqref{e:inv} using geometric desingularization. We think of $(W,V,\eta)\in [0,\infty)\times S^2$, using polar coordinates, thus blowing up the origin into a ``singular'' sphere. The dynamics on the sphere are then analyzed using stereographic projections. The singular sphere of course should be identified with the origin, a single equilibrium and therefore consists of equilibria, only. The key idea is to use nonlinear rescalings of time such that this sphere  exhibits a nontrivial flow which allows one to conclude behavior in a vicinity of the origin; see \cite{dum,ks} for illustrations and applications of this geometric desingularization technique. 

Before diving into this construction, we emphasize that the literature here uses the terminology ``geometric blowup'', referring to the fact that a single point, the origin, is blown up into a sphere. This ``blowup'' has no relation to the fact that we are interested in self-similar solutions to a PDE that eventually blow up in finite time!

We start by identifying the natural scaling $V\sim W\sim \eta^{1/2}$, with leading order equation,
\begin{align}
 {W}_\sigma&=2W^2(-\frac{1}{2} W+V),\notag\\
 {V}_\sigma&=-\eta V,\notag\\
 {\eta}_\sigma&=-2\eta W^2.\label{e:invl}
\end{align}
We next introduce three coordinate charts that describe the flow in the neighborhood of the origin and correspond to stereographic projections of the flow in polar coordinates,
\begin{equation}
\begin{array}{lll} 
\eta\mbox{-directional blowup:}\qquad& V_1=V/\eta^{1/2},& W_1=W/\eta^{1/2},\notag\\[0.1in]
 W\mbox{-directional blowup:}\qquad& V_2=V/W,& \eta_2=\eta/W^2,\notag\\[0.1in]
 V\mbox{-directional blowup:}\qquad& W_3=W/V,& \eta_3=\eta/V^2.
\end{array}\label{e:bu}
\end{equation}
We next compute the resulting equations. 

\textbf{$\eta$-directional blowup.} We use the Euler multiplier $\partial_\tau=\eta^{-1}\partial_\sigma$ and find at leading order in $\eta$, 
\begin{align}
 V_{1,\tau}&=V_1(W_1^2-1)+\rmO(\eta^{1/2}),\notag\\
 W_{1,\tau}&=V_1W_1^2+\rmO(\eta^{1/2}),\notag\\
 \eta_\tau&=-2\eta W_1^2+\rmO(\eta^{3/2}).\label{e:1ch}
\end{align}
The flow in $\eta=0$ can be readily analyzed exploiting the Euler multiplier $V_1$ and explicitly integrating 
\[
 V_{1,s}=W_1^2-1,\qquad W_{1,s}=W_1^2,
\]
to yield the flow depicted in Figure \ref{f:charts}, top left. Note in particular that solutions with $V_1,W_1>0$ will leave a neighborhood of the origin in this chart, that is, there are no nontrivial solutions that remain bounded in this chart for all positive  times. 

\textbf{$W$-directional blowup.} We use the Euler multiplier $W^2$ to arrive at 
\begin{align}
 \eta_{2,\tau}&=-4\eta_2V_2+\rmO(W),\notag\\
 V_{2,\tau}&=V_2(1-\eta_2-2V_2)+\rmO(W),\notag\\
 W_\tau&=2W(V_2-\frac{1}{2})+\rmO(W^2).\label{e:2ch}
\end{align}
The flow in the singular sphere $W=0$ can again be analyzed explicitly using the Euler multiplier $V_2$, which yields 
\[
 \eta_{2,s}=-4\eta_2,\qquad V_{2,s}=1-\eta_2-V_2. 
\]
The resulting flow is depicted in Figure \ref{f:charts}, top right.
\begin{figure}[h]
 \centering\includegraphics[width=0.65\textwidth]{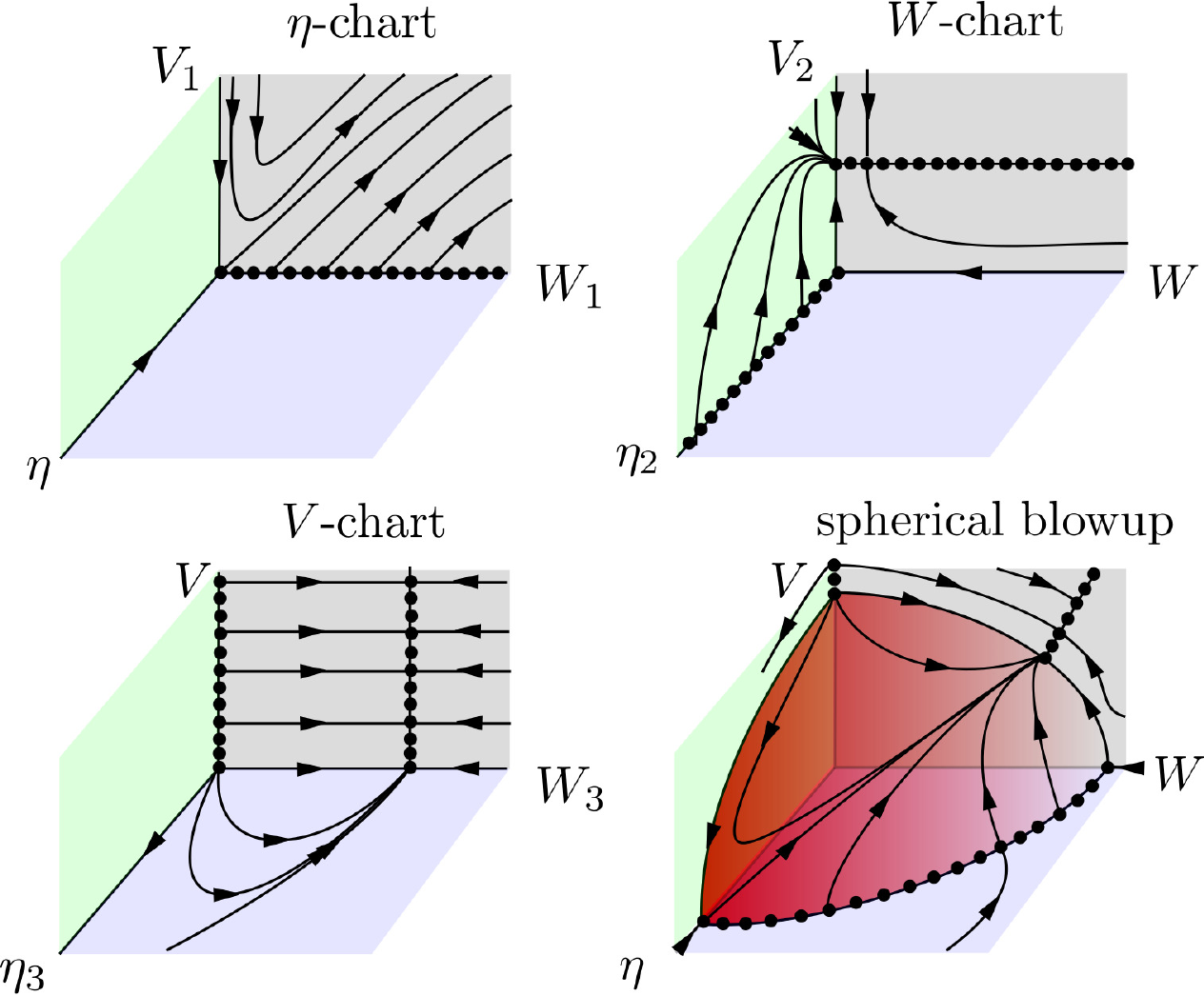}
 \caption{Flows to \eqref{e:1ch} (top left),\eqref{e:2ch} (top right), and \eqref{e:3ch} (bottom left), and reconstructed flow on the singular sphere with normal dynamics (bottom right). In the individual charts, the singular sphere is given by the coordinate planes $\eta=0$, $W=0$, $V=0$ (gray, green, blue), respectively; in the last assembled spherical blowup, the singular sphere is shaded red. The individual charts show the flow in the last picture near the associated coordinate axes. }\label{f:charts}
\end{figure}

\textbf{$V$-directional blowup.} We use the Euler multiplier $W^2$ to arrive at
\begin{align}
 W_{3,\tau}&=W_3(\eta_3+2W_3(1-\frac{1}{2}W_3))+\rmO(V),\notag\\
 \eta_{3,\tau}&=\eta_3(2\eta_3-2W_3^2)+\rmO(V),\notag\\
 V_\tau&=-V\eta_3+\rmO(V^2).\label{e:3ch}
\end{align}
The flow can again be analyzed using elementary methods to yield the phase portrait depicted in Figure \ref{f:charts}, bottom left. Undoing the stereographic projections, we arrive at the flow on the positive octant as depicted in Figure \ref{f:charts}, bottom right. \footnote{Note that the strong stable manifold of the nontrivial equilibria $W_3=2,\eta_3=0$ converges in backward time to the origin  $\eta_3=V_3=0$ along the stable ray solution $\eta_3=2W_3-W_3^2$, as is most easily seen exploiting the explicit representation of the strong stable manifold in the $W$-chart and converting back.} Having characterized the flow in a vicinity of the singular sphere, we are now ready to conclude the proof. 

From the phase portrait, we recognize that the $\omega$-limit set of any trajectory is contained in the boundary of the positive octant. The stable manifolds of the equilibria on the coordinate axes, and the equilibria in $V=0$ are contained in the coordinate axes and in $V=0$, respectively. All other trajectories converge to the family of equilibria limiting on $V_2=1/2$, $\eta_2=0$. The stable manifold of the singular equilibrium is contained entirely in the singular sphere, by uniqueness, such that all trajectories limit on an equilibrium with $W>0$, hence proving that $U=1/W$ stays bounded as claimed. 
\end{Proof}

\section{Global existence and bounds: $\eps>0$}\label{s:6}
In this section, we focus on 
\begin{align}
u_t&=u_{xx}+u_x-ug(v)+vg(u),\\
v_t&=v_{xx}-v_x+ug(v)-vg(u), \label{parabolic-10}
\end{align}
with periodic boundary conditions on $x\in(0,L)$, and assume $g(0)\geq 0$ to guarantee that positivity of solutions is preserved. 

\begin{Theorem}\label{t:dgl}
 Suppose $g\in C^1$ and consider initial conditions $0\leq u_0(x),v_0(x)\in L^\infty$. Then the unique solution to \eqref{parabolic-10} exist for all $t>0$ and 
 \[
  \limsup_{t\rightarrow\infty}\norm{w(t)}_{{\infty}}\leq C(\norm{u_0}_\infty,\norm{v_0}_\infty).
 \]
\end{Theorem}

\begin{Proof}
Setting $w=u+v$, we obtain
\[
w_t-w_{xx}-u_x+v_x=0. 
\]
Multiplying by $w^{q-1}$, for any $q>1$, we obtain the a priori estimate 
\begin{eqnarray*}
\frac{1}{q}\frac{\rmd}{\rmd t}\int
w^q+\frac{4(q-1)}{q^2}\int \abs{\partial_x
w^{\frac{q}{2}}}^2&=& \int(u_x-v_x)w^{q-1}
= -\int(u-v) w^{q-1}_x\\
&\leq&\frac{2(q-1)}{q}\int\abs{\partial_x
w^{\frac{q}{2}}}w^{\frac{q}{2}}
\leq \frac{2(q-1)}{q^2}\int
\abs{\partial_x w^{\frac{q}{2}}}^2+\frac{q-1}{2}\int w^q,
\end{eqnarray*}
which implies
\begin{equation}\label{KSS-5}
\frac{1}{q}\frac{\rmd}{\rmd t}\int
w^q+\frac{2(q-1)}{q^2}\int \abs{\partial_x
w^{\frac{q}{2}}}^2\le \frac{q-1}{2}\int w^q,
\end{equation}
and, using Gronwall's inequality,
\[
\norm{w(t)}_{L^p}\le \norm{w_0}_{L^p}\exp\bke{\frac{p-1}{2}t},\qquad
1\le p<\infty.
\]
Since $u,v\geq 0$, this readily gives global existence and bounds on $(u,v)$. 
Next, we show that, in fact, $(u, v)$ are uniformly bounded in time. 

We use \eqref{KSS-5} with $q=2$ and find 
\begin{equation}\label{KSS-51}
\frac{\rmd}{\rmd t}\int
w^2+\int \abs{\partial_x w}^2\le \int w^2.
\end{equation}
We exploit that fact that $\|w\|_1\equiv\|w_0\|_1$ is conserved in time via a Gagliardo-Nirenberg inequality,
\begin{equation}\label{e:gn}
\norm{f}_{L^2}\le
C\norm{f}^{\frac{2}{3}}_{L^1}\norm{\partial_x
f}^{\frac{1}{3}}_{L^2}+C\norm{f}_{L^1},
\end{equation}
which readily gives 
\begin{equation}\label{e:gn2}
\norm{\partial_x
w}_{L^2}^2\geq c_1\frac{\norm{w}_{L^2}^6}{\norm{w}_{L^1}^4}-c_2\norm{w}_{L^1}^4.
\end{equation}
Substituting into \eqref{KSS-51} results in a differential inequality for $\norm{w}_{L^2}^2=:y(t)$, 
\[
y'(t)\le C_1+y(t)-C_2y^{3}(t),\qquad 
y(0)>0.
\]
Comparing with the solution to the differential equation, we immediately find  $\displaystyle\limsup_{t\rightarrow\infty}
\norm{w(t)}_{L^2(\R^d)}\le C(\|w_0\|_1)$.

Next, we derive bounds on 
$\displaystyle\limsup_{t\rightarrow\infty}\norm{w(t)}_{L^q(\R^d)}$,
$q=2^k$, where $k$ is a positive integer. We suppose that
$\norm{w(t)}_{L^{2^m}(\R^d)}$, $m=k-1$ is uniformly bounded by a
constant $M_{k-1}$ as $t \rightarrow \infty$, that is
$\limsup_{t\rightarrow\infty}\norm{w(t)}_{L^{2^m}}:=M_{k-1}$.
Taking $\tilde{w}=w^{\frac{q}{2}}$ with $q=2^k$, we obtain from
\eqref{KSS-5} that
\[
\frac{\rmd}{\rmd t}\norm{\tilde{w}}^{2}_{L^2}+\frac{2(2^k-1)}{2^k}\norm{\partial_x
\tilde{w}}^{2}_{L^2}\le
2^k(2^k-1)\norm{\tilde{w}}^{2}_{L^2},
\]
which gives, using \eqref{e:gn2},
\[
\frac{\rmd}{\rmd t}\norm{\tilde{w}}^{2}_{L^2}+c_1\frac{2^k-1}{2^{k-1}}
\frac{\norm{\tilde{w}}^{6}_{L^2}}{\norm{\tilde{w}}^{4}_{L^1}}-c_2\frac{2^k-1}{2^{k-1}}\norm{\tilde{w}}^{2}_{L^1}
\le
2^k(2^k-1)\norm{\tilde{w}}^{2}_{L^2},
\]
that is,
\[
\frac{\rmd}{\rmd t}\norm{\tilde{w}}^{2}_{L^2}\le
(2^k-1)\left(\norm{\tilde{w}}^{2}_{L^2}\bke{2^k-c_1\frac{2^k-1}{2^{k-1}}
\frac{\norm{\tilde{w}}^{4}_{L^2}}{\norm{\tilde{w}}^{4}_{L^1}}}+c_2\frac{1}{2^{k-1}}\norm{\tilde{w}}^{2}_{L^1}\right).
\]
Note that the right-hand side of this differential inequality is negative when
\[
 \norm{\tilde{w}}_{L^2}\leq \frac{2^k2^{k-1}}{{c_3}(2^k-1)}\norm{\tilde{w}}_{L^1},
\]
for some ${c}_3$. 
Since
$\displaystyle\limsup_{t\rightarrow\infty}\norm{\tilde{w}}_{L^1}
=\displaystyle\limsup_{t\rightarrow\infty}\norm{w}^{2^{k-1}}_{L^{2^{k-1}}}=
M_{k-1}^{2^{k-1}}$, we can see that
\[
\limsup_{t\rightarrow\infty}\norm{w}^{2^k}_{L^{2^k}}=\limsup_{t\rightarrow\infty}\norm{\tilde{w}}^2_{L^2}\le
\bke{\frac{2^k2^{k-1}}{c_3(2^k-1)}}^{\frac{1}{2}}\norm{\tilde{w}}^2_{L^1}\le
\bke{\frac{2^k2^{k-1}}{c_3(2^k-1)}}^{\frac{1}{2}}M_{k-1}^{2^{k}},
\]
which yields
\[
M_{k}=\limsup_{t\rightarrow\infty}\norm{w}_{L^{2^k}}\le
\bke{\frac{2^k}{c_3}}^{\frac{d}{2^{k+1}}}M_{k-1}.
\]
This iteration formula shows that $M_k$ are all bounded for each $k$
and furthermore, $\displaystyle\limsup_{k\rightarrow\infty} M_k\le
2M_0$, which implies that
$\limsup_{t\rightarrow\infty}\norm{w(t)}_{L^{\infty}}$ is also
uniformly bounded.
\end{Proof}

\section{Discussion}
We presented a number of results that aim at deducing coarse qualitative properties of the possibly complex dynamics of run-and-tumble system in terms of simple properties of asymmetric, spatially homogeneous equilibria. Coarse qualitative properties, as defined here, are convergence to symmetric, spatially homogeneous equilibria, termed equidistribution, existence of wave patterns, and, lastly, the possibility of blowup. 

In this regard, our results provide fairly comprehensive information, with a few notable exceptions. First, in the absence of asymmetric equilibria, we establish equidistribution only for $\eps>0$, or under additional assumptions. It is an interesting question if one can relax these assumptions and establish convergence more generally. In the regime of patterns and waves, $\Upsilon_\mathrm{I}$, many questions are open in regard to a description of the dynamics. In the absence of diffusion, selection of wavenumbers through shot noise perturbations was studied analytically and numerically in \cite{scst}. In the presence of diffusion, fronts and waves have been analyzed in \cite{fns,fuhrmann}. Stability of waves, in particular in the absence of diffusion, appears to be a realistic first step towards a better understanding of the plethora of wave patterns that exist in $\Upsilon_\mathrm{I}$. Lastly, in the regime $\Gamma_\infty$, $\eps=0$, one would wish to find more robust results on blowup, pertaining for instance to the stability of blowup profiles, the existence of blowup for $\mu>0$, that is, in the absence of scale invariance, or for different values of $p,q$ in \eqref{e:g}. 

Turning back to the applications and critical turning rates discussed in \cite{Lutscher-Stevens}, which motivated \cite{scst}, our analysis suggests simple switching mechanism, where small changes of parameters lead to changes in the collective, self-organized behavior of population, changing between equidistribution, rippling, and the formation of clusters with potential blowup, thus relating to experimentally observed formation of rippling
patterns and the onset of fruiting body formation due to changes in food supply.

While our analysis suggests that such simple parameter changes may be sufficient, it is of course by no means clear that the parameter changes described here are in fact major contributors to the phase transitions observed in experiments. Beyond the inclusion of further terms in the analysis, modeling for instance delays in turning events, a more direct extension would be towards two-dimensional domains, with velocities distributed on the unit-circle $S^1$ rather than in $S^0$. It would be interesting to examine equidistribution, existence of aligned ripples, and possible blowup in this more general scenario.


\begin{thebibliography}{20}
\bibitem{attr} A.V. Babin and M.I. Vishik.
\textit{Attractors of evolution equations. }
Translated and revised from the 1989 Russian original by Babin. Studies in Mathematics and its Applications, \textbf{25}. North-Holland Publishing Co., Amsterdam, 1992.
\bibitem{Childress-Percus} S. Childress and J.K. Percus.
\textit{Nonlinear aspects of chemotaxis.}
Math. Biosci. \textbf{56} (1981), 217--237.
\bibitem{dum}
F. Dumortier. \textit{Techniques in the theory of local bifurcations: blow-up, normal forms, nilpotent bifurcations, singular perturbations. Bifurcations and periodic orbits of vector fields} (Montreal, PQ, 1992), 19--73, NATO Adv. Sci. Inst. Ser. C Math. Phys. Sci., \textbf{408}, Kluwer Acad. Publ., Dordrecht, 1993. 
\bibitem{fns} P. Flynn, Q. Neville, and A. Scheel. 
\textit{Self-organized clusters in diffusive run-and-tumble processes.}
Submitted.
\bibitem{Freistuehler-Fuhrmann} H. Freist\"uhler and J. Fuhrmann.
\textit{Traveling waves in directed diffusive particle flows.}
SIAM J. Appl. Math. \textbf{78} (2018),  759--773.
\bibitem{fuhrmann} J. Fuhrmann. 
\newblock On a Minimal Model for the Initiation of Cell Movement.
\newblock \textit{Dissertation, Universit\"at Heidelberg}, 2012; DOI: 10.11588/heidok.00013659.
\bibitem{attr2} J. Hale. 
\textit{Asymptotic behavior of dissipative systems.}
Mathematical Surveys and Monographs, \textbf{25}. American Mathematical Society, Providence, RI, 1988.
\bibitem{Hillen97} T. Hillen. 
\textit{Invariance principles for hyperbolic random walk systems.}
J. Math. Anal. Appl. \textbf{210} (1997), 360--374.
\bibitem{Jaeger-Luckhaus} W. J\"ager and S. Luckhaus.
\textit{On explosions of solutions to a system of partial differential equations
modelling chemotaxis.}
Trans. Amer. Math. Soc. \textbf{329} (1992),  819--824.
\bibitem{ks} M. Krupa and P. Szmolyan.
\textit{Extending geometric singular perturbation theory to nonhyperbolic points—fold and canard points in two dimensions.}
SIAM J. Math. Anal. \textbf{33} (2001),  286--314. 
\bibitem{lun} A. Lunardi.
\textit{Analytic semigroups and optimal regularity in parabolic problems. }  Birkh\"auser/Springer Basel AG, Basel, 1995.
\bibitem{Lutscher} F. Lutscher.
\textit{Modeling alignment and movement of animals and cells.}
J. Math. Biol.  \textbf{45} (2002) 234--260.
\bibitem{Lutscher-Stevens} F. Lutscher and A. Stevens.
\textit{Emerging patterns in a hyperbolic model for locally interacting
cell systems.}
J. Nonlinear Sci. \textbf{12} (2002), 619--640.
\bibitem{Nanjundiah} V. Nanjundiah.
\textit{Chemotaxis, signal relaying and aggregation morphology.}
J. Theor. Biol. \textbf{42} (1973),  63--105.
\bibitem{Reich} H. Reichenbach.
\textit{Rhythmische Vorgänge bei der Schwarmentfaltung von Myxobakterien.}
Ber. Deutsch. Bot. Ges. \textbf{78} (1965), 102.
\bibitem{scst} A. Scheel and  A. Stevens.
\textit{Wavenumber selection in coupled transport equations.}
J. Math. Biol. \textbf{75} (2017), 1047--1073.
\bibitem{ShKai} L.J. Schimkets and D. Kaiser.
\textit{Induction of Coordinated Movement of Myxococcus xanthus.}
J. of Bacteriol. \textbf{152} (1982), 451--461.
\bibitem{smoller} J. Smoller.
\textit{Shock waves and reaction-diffusion equations.} 
Grundlehren der Mathematischen Wissenschaften, \textbf{258}. Springer-Verlag, New York-Berlin, 1983.
\end{thebibliography}
\end{document}